\documentclass[aps, superscriptaddress,reprint,amsmath,amssymb,prl]{revtex4-2}
\usepackage{graphicx}
\graphicspath{{../figures/}}
\usepackage{multirow}
\usepackage{dcolumn}
\usepackage{float}
\usepackage[usenames]{color}
\usepackage{soul} 
\usepackage{booktabs}
\usepackage{siunitx}
\usepackage{amsmath}
\usepackage[version=4]{mhchem}
\usepackage[colorlinks=true, allcolors=blue]{hyperref}
\usepackage{makecell}

\begin{document}

\title{How Atoms Interact Within Molecules}
\author{Adil Kabylda}
\affiliation{Department of Physics and Materials Science, University of Luxembourg, L-1511 Luxembourg City, Luxembourg}
\author{Malte Esders}
\affiliation{Machine Learning Group, Technische Universit\"at Berlin, 10587 Berlin, Germany}
\affiliation{Berlin Institute for the Foundations of Learning and Data -- BIFOLD, 10587 Berlin, Germany}
\author{Matteo Gori}
\affiliation{Department of Physics and Materials Science, University of Luxembourg, L-1511 Luxembourg City, Luxembourg}
\author{Stefan Chmiela}
\affiliation{Machine Learning Group, Technische Universit\"at Berlin, 10587 Berlin, Germany}
\affiliation{Berlin Institute for the Foundations of Learning and Data -- BIFOLD, 10587 Berlin, Germany}
\author{Klaus-Robert M\"{u}ller}
\email{klaus-robert.mueller@tu-berlin.de}
\affiliation{Machine Learning Group, Technische Universit\"at Berlin, 10587 Berlin, Germany}
\affiliation{Berlin Institute for the Foundations of Learning and Data -- BIFOLD, 10587 Berlin, Germany}
\affiliation{Max Planck Institute for Informatics, Stuhlsatzenhausweg, 66123 Saarbr{\"u}cken, Germany}
\affiliation{Department of Artificial Intelligence, Korea University, Anam-dong, Seongbuk-gu, Seoul 02841, Korea}
\author{Alexandre Tkatchenko}
\email{alexandre.tkatchenko@uni.lu}
\affiliation{Department of Physics and Materials Science, University of Luxembourg, L-1511 Luxembourg City, Luxembourg}

\begin{abstract}
\noindent 
Fundamental understanding of interatomic forces in molecules must emerge from quantum mechanics, yet widely used empirical force fields rely on simplified mechanistic approximations that often fail to capture the complexity of many-body systems. Here we employ recent developments in quantum field theory (QFT) for long-range electron correlation and machine learning force fields (MLFFs) to directly compute the depth and scatter of interatomic forces for molecular systems containing hundreds of atoms. We find that while the average interaction strength decays polynomially with interatomic separation, the interaction scatter remains robust and exhibits substantial anisotropy. Both QFT and MLFFs demonstrate that increasing the molecular size further amplifies this scatter and anisotropy -- a phenomenon not considered in traditional textbook empirical models. These results provide new benchmarks for force models, shift the focus from interacting atoms to interacting ``hotspots'' that might determine the folding pathways of (bio)polymers, and rationalize why MLFFs are uniquely successful in capturing the nuances of complex molecular systems. Our findings offer a roadmap for the development of more accurate and quantum-aware molecular force fields.
\end{abstract}

\maketitle

\section{Introduction}
\vspace{-2mm}

Understanding the nature of interatomic interactions is fundamental to molecular and materials science, as it directly impacts our ability to model and predict their behavior. Although all interactions are inherently quantum mechanical, their complexity has historically necessitated the use of simplified models~\cite{frenkel2023understanding, cornell1995second, mackerell1998all}. The characteristics of interatomic forces, particularly their range, anisotropy (angular deviation from the vector connecting two atoms), and functional dependence, remain an area of active investigation~\cite{stone2013theory, hermann2017first, stone2007atom}. The central questions are: What is the effective distance range of interatomic interactions needed for accurate simulations, how does it scale with molecular size, and how anisotropic are interactions at various interatomic distances?

\begin{figure*}
    \centering
    \includegraphics[width=.99\linewidth]{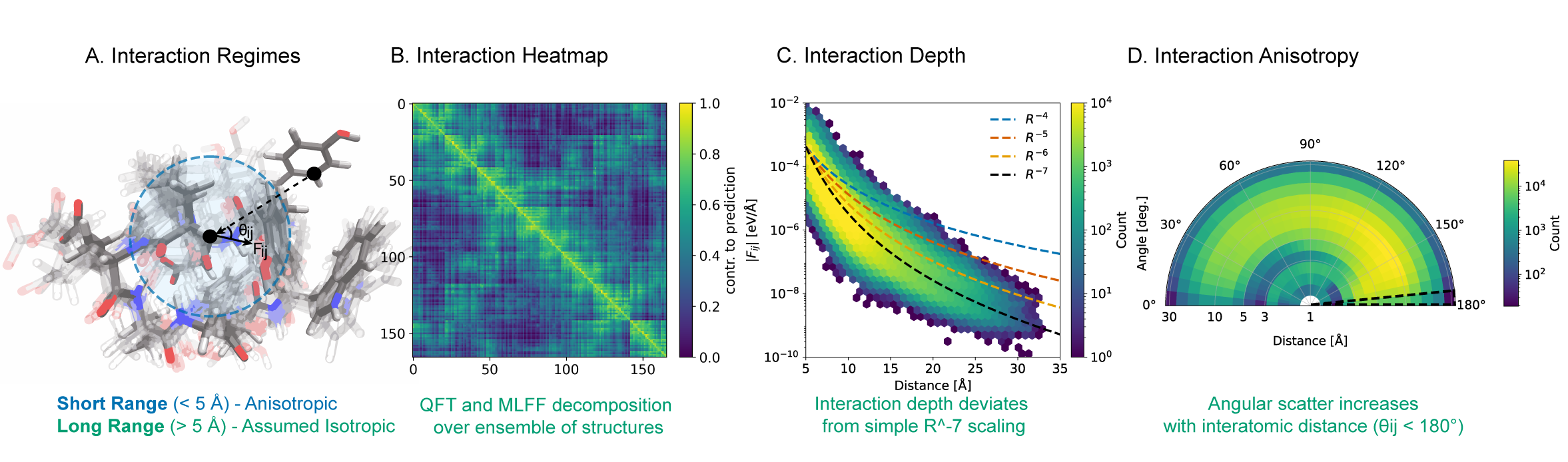}
    \caption{\textbf{Long-range electron correlation interactions exhibit non-uniform decay and growing anisotropy in a protein.} \textbf{(A)} Framework for analyzing force contributions $F_{ij}$ between atoms $i$ and $j$, where $\sum_j F_{ij} = F_i$ gives the total force on atom $i$. The angle $\theta_{ij}$ measures the alignment between the interatomic vector $d_{ij}$ and the force contribution $F_{ij}$. Short-range interactions ($< 5$~\text{Å}) are known to exhibit anisotropy, while long-range interactions are often assumed isotropic in force fields.
    \textbf{(B)} Interaction heatmap for Chignolin ($F_{ij}$), obtained through QFT and MLFF decompositions averaged over 100 snapshots from an \emph{ab initio} molecular dynamics trajectory  spanning folded, semi-folded, and unfolded conformations (see Methods for details).
    \textbf{(C)} Interaction depth, i.e. the decay of the interaction strength $|F_{ij}|$ with interatomic distance. The dashed $R^{-4}$ to $R^{-7}$ reference curves are anchored at $R = 5$~\AA{} to a global fit of $|F_{ij}| = c\,R^{-7}$ over the $5$--$10$~\text{Å} range. The 2D histogram shows substantial scatter at each distance with non-uniform decay.
    \textbf{(D)} Polar representation of the angular $\theta_{ij}$ distribution vs. interatomic distance. The heatmap demonstrates that directional dependence (deviation from $\theta_{ij} = 180^\circ$, which corresponds to pairwise attractive interactions; dashed sector) grows systematically at longer range. The magnitude and directional character of long-range forces exhibit complex, environment- and distance-dependent behavior not captured by pairwise potentials.
    }
    \label{fig:overview}
\end{figure*}

The motivation for this study goes back to the pioneering work of Langmuir and Gurvich more than a century ago, who debated the characteristic distances of interatomic and intermolecular interactions, proposing ranges from roughly 1 to 300~\text{Å}~\cite{gurwitsch1914physiko, langmuir1917constitution}. Today, advances in computational methods and experimental techniques have refined these scales, suggesting that critical bonded interactions occur below 5~\text{Å} distances (Fig.~\ref{fig:overview}A). Although short-range anisotropy in interatomic potentials has been well-established (for example, due to non-spherical valence electron distributions such as lone pairs and $\pi$-bonds)~\cite{stone1988some}, long-range anisotropy has received comparatively little attention. Traditional models tend to assume isotropy at longer ranges, relying on pairwise additive schemes for non-covalent electrostatic and dispersion forces that decay as $R^{-2}$ or $R^{-7}$, respectively~\cite{jones1924determination, london1937general, tkatchenko2009accurate, grimme2010consistent}.

However, these assumptions neglect the multiscale nature of interactions that potentially leads to anisotropic contributions that may persist or amplify at greater distances between atoms or larger molecular fragments. Indeed, evidence from multiple domains of molecular science points to a more complex picture. Second derivatives of the \textit{ab initio} energy surface~\cite{dinur1989direct, hauseux2020quantum, hauseux2022colossal}, orientation-dependent fits from high-level quantum theory~\cite{van2018new}, and symmetry-adapted perturbation theory (SAPT) studies quantifying anisotropy in exchange and dispersion~\cite{szalewicz2012symmetry, kriz2024quantification}, all show that the standard ``sum-of-spheres'' approximation fails unless non-spherical features such as virtual sites or angular terms are explicitly included. Experimentally, phenomena such as rapid specific protein association~\cite{schreiber1996rapid} and the anisotropic inter-layer van der Waals potentials of 2D materials point to long-range forces acting as directional ``steering'' mechanisms rather than isotropic attractions~\cite{stone1988some, price2000toward}. Given that such persistent anisotropy is evident even in small simple molecules, the assumption that it averages out or vanishes in larger, more complex systems (in particular, for atoms interacting \textit{within} molecules) is not well founded.

Beyond the covalent distance regime, there are several kinds of contributions to interatomic interactions. They can arise from permanent, induced, and fluctuating multipoles. Electrostatic and polarization interactions depend on permanent and induced dipoles and are usually screened in extended systems~\cite{honig1995classical, ren2012biomolecular}, either because multipoles do not scale with molecular size or due to orientational polarization and multipole cancellation~\cite{stone2013theory, zhou2018electrostatic}. Dispersion interactions involve both fluctuating and induced multipoles and depend on high-frequency (not static) dielectric properties of matter. Dispersion interactions are usually attractive between ground-state fragments, scale with system size, and arise from correlated zero-point plasmonic modes in the optical frequency range that are wave-like and delocalized throughout the molecular framework~\cite{ambrosetti2016wavelike, stohr2019quantum, gori2023second}. Consequently, dispersion forces can converge slower with distance than electrostatic forces (see Fig.~\ref{fig:elec_mbd_conv} for molecules studied in this work).

This study examines how atoms interact within molecules by leveraging new developments in Quantum Field Theory (QFT) for long-range electron correlation~\cite{karimpour2026quantum} and Machine Learning Force Fields (MLFF)~\cite{unke2021machine}. These methods allow us to probe how interaction depth and directional anisotropy evolve across length scales and system sizes. Across systems ranging from small biomolecular units to a 35-residue protein, we find that the interaction depth exhibits broad scatter that defies simple power-law decay, and that the anisotropy of interatomic forces persists at long-range and amplifies with system size.

\section{Results}
\vspace{-2mm}

The central challenge in characterizing interatomic interactions is disentangling the contributions of individual atom pairs from the collective quantum mechanical behavior of the full many-body molecular system. To address this, we develop and apply a framework for extracting pairwise force contributions $F_{ij}$ and their angular alignments $\theta_{ij}$ from three distinct computational approaches: Second-Quantized Many-Body Dispersion (SQ-MBD) theory~\cite{tkatchenko2012accurate, distasio2014many, gori2023second} (with and without mean-field DFT contributions), the PaiNN machine learning force field~\cite{schutt2021equivariant}, and a Mechanistic Empirical Force Field (MEFF). These methods span a wide range of physical approximations, from the fully quantum, non-local description of SQ-MBD to the atom-type-based MEFF, enabling a systematic assessment of which physical ingredients are necessary to reproduce the true force landscape.

We apply this framework to five molecular systems of diverse structural complexity: the tetrapeptide \ce{AcAla3NMe} (42 atoms), the fatty acid DHA (56 atoms), the supramolecular Buckyball Catcher complex (148 atoms)~\cite{chmiela2023accurate}, the Chignolin miniprotein (166 atoms)~\cite{wang2023aimd}, and the FIP35 WW domain (562 atoms)~\cite{stohr2019quantum, lindorff2011fast}. The two proteins bracket a biologically relevant size range: Chignolin is a 10-residue $\beta$-hairpin already containing aromatic, polar, and conformationally compact motifs typical of larger proteins~\cite{honda200410}, while FIP35 extends the analysis to a system with a full tertiary structure~\cite{piana2011computational}. 

When interaction strength $|F_{ij}|$ is plotted against interatomic distance for Chignolin mini-protein (Fig.~\ref{fig:overview}C), the SQ-MBD data do not collapse onto a single power-law curve as conventional $R^{-7}$ dispersion force models would predict. Instead, a broad cloud of interaction strengths spanning up to four orders of magnitude emerges at every distance. We further quantify the directional character through the alignment angle $\theta_{ij}$ between $F_{ij}$ and the interatomic vector $d_{ij}$. In a purely pairwise-additive attraction $\theta_{ij} = 180^\circ$, whereas deviations signal many-body anisotropy. The angular distribution (Fig.~\ref{fig:overview}D) shows that this directional dependence grows systematically with interatomic separation — the fraction of forces aligned along the interatomic axis ($\theta_{ij} > 150^\circ$) drops from roughly 50\% at short range to below 10\% beyond 15~\text{\AA}, with the remaining forces distributed over a broad angular range. The observed variability implies that the spatial distribution of interacting atoms and fragments creates a force landscape that is substantially more complex than previously appreciated.

In the following, we first outline the origin of force scatter and anisotropy through a dimensional analysis of quantum mechanics, then introduce the SQ-MBD formalism and define the pairwise force decomposition $F_{ij}$ that is analyzed throughout (Fig.~\ref{fig:overview}B; full derivations for SQ-MBD, MLFF, and MEFF are provided in the Methods). We next present a quantitative examination of interaction anisotropy and interaction depth (the decay of interaction strength with interatomic distance) across the molecular test set, and finally identify the residue-level hotspots and their dependence on the folding state.

\subsection{Dimensional structure of quantum mechanics and the origin of interaction scatter}

The existence of force scatter and anisotropy can be understood through a dimensional analysis that traces the projection from molecular structure to interatomic forces. A molecular system of $N$ atoms is fully specified by its nuclear charges $\{Z_i\}$ and positions $\{R_i\} \in \mathbb{R}^{3N}$, a representation embedded in a $4N$-dimensional space that uniquely determines the electronic Hamiltonian (in atomic units)
\begin{equation}
\label{eq:full_H}
\hat{H} =
    -\sum_i \frac{\nabla_i^2}{2}
    - \sum_{i,A} \frac{Z_A}{|\mathbf{r}_i - \mathbf{R}_A|}
    + \sum_{i<j} \frac{1}{|\mathbf{r}_i - \mathbf{r}_j|}.
\end{equation}

By solving the Schr\"{o}dinger equation $\hat{H}|\Psi\rangle = E|\Psi\rangle$, this finite specification generates a ground-state wavefunction $|\Psi\rangle$ embedded in the $L^2$-Hilbert space $\mathcal{H}$ of $N_e$-electron states, a space of formally infinite dimension. The interatomic forces are then recovered as expectation values via the Hellmann-Feynman theorem,
\begin{equation}
\label{eq:HF}
\mathbf{F}_i = -\frac{\partial E}{\partial \mathbf{R}_i}
    = -\bigl\langle \Psi \bigl| \frac{\partial \hat{H}}{\partial \mathbf{R}_i} \bigr| \Psi \bigr\rangle,
\end{equation}
collapsing the quantum state back onto a $3N$-dimensional vector. The full sequence is therefore a map between spaces of strikingly different dimension,
\begin{equation}
\label{eq:dim_chain}
\underbrace{\{Z_i, \mathbf{R}_i\}}_{\dim\, =\, 4N}
\;\xrightarrow{\;\hat{H}\;}
\underbrace{|\Psi\rangle \in \mathcal{H}}_{\dim\, =\, \infty}
\;\xrightarrow{\;\langle \partial_{R} \hat{H} \rangle\;}
\underbrace{\{\mathbf{F}_i\}}_{\dim\, =\, 3N}
\;\xrightarrow{\;\text{SQ}\;}
\underbrace{\{\mathbf{F}_{ij}\}}_{\dim\, =\, N \times 3N}
\end{equation}
a passage through an infinite-dimensional space that connects two finite-dimensional projections of it.

This dimensional structure makes clear why pairwise distances alone are poor predictors of interatomic forces. Molecular structure is specified compactly: $4N$ numbers determine the Hamiltonian, which in turn determines a complex wavefunction embedded in an infinite dimensional space, from which $3N$ forces are finally extracted. The forces carry the imprint of the full quantum mechanical complexity. Two atom pairs sitting at the same separation $R_{ij}$ but embedded in different molecular environments will generally experience different forces, because the wavefunction encodes the positions and identities of all other atoms simultaneously. The resulting variability in $|F_{ij}|$ at fixed distance is not a statistical fluctuation but a physically meaningful signal, and it grows with system size as the number of atoms shaping the wavefunction increases. This is why simple cutoff schemes and pairwise additive force fields become less reliable as molecules grow in complexity. We now describe how SQ-MBD formalizes this picture to yield an effective, computable pairwise force decomposition for the long-range electron correlation energy (the most non-trivial quantum-mechanical part of the electronic energy).

\subsection{Second quantization of many-body dispersion}

Long-range electron correlation is a fundamental part of the electronic energy, and it arises from correlated quantum fluctuations in the electronic density~\cite{hermann2017first}. The many-body dispersion (MBD) formalism~\cite{tkatchenko2012accurate, distasio2014many} captures these contributions by representing each atom as a quantum harmonic oscillator carrying an electric dipole, known as a Quantum Drude Oscillator (QDO)~\cite{jones2013quantum, khabibrakhmanov2025accurate}, parameterized to reproduce the electric-response properties of atoms in molecules. When isolated, each QDO fluctuates independently; once coupled through dipole--dipole interactions, the oscillators develop delocalized collective modes in which all atoms participate simultaneously. It is the correlated zero-point motion of these collective modes that gives rise to the many-body dispersion energy.

Because the QDO Hamiltonian is quadratic in the displacement and momentum operators, it can be exactly diagonalized into these collective modes. In the second-quantization framework~\cite{gori2023second}, it takes the form
\begin{equation}
\label{eq:MBD_H}
\hat{H}_{\mathrm{SQ\text{-}MBD}} = \sum_{k=1}^{3N} \hbar\tilde{\omega}_k
\left(\hat{b}_k^\dagger\,\hat{b}_k + \tfrac{1}{2}\right),
\end{equation}
where $\tilde{\omega}_k$ are the eigenfrequencies of the collective MBD modes and $\hat{b}_k^\dagger, \hat{b}_k$ are the corresponding bosonic creation and annihilation operators. The connection between these collective operators and the ladder operators $\hat{a}_{i}^\dagger, \hat{a}_{i}$ of the individual atomic QDOs is given by a Bogoliubov transformation~\cite{berazin2012method, blaizot1986quantum},
\begin{equation}
\label{eq:bogoliubov}
\begin{pmatrix} \hat{\boldsymbol{b}} \\[2pt] \hat{\boldsymbol{b}}^\dagger \end{pmatrix}
=
\begin{pmatrix} X & Y \\[2pt] Y & X \end{pmatrix}
\begin{pmatrix} \hat{\boldsymbol{a}} \\[2pt] \hat{\boldsymbol{a}}^\dagger \end{pmatrix},
\end{equation}
where the real matrices $X$ and $Y$ are determined by the orthogonal transformation that diagonalizes the MBD potential matrix together with the atomic and collective eigenfrequencies. This transformation encodes the information content of the MBD ground state $|\tilde{0}\rangle$ in terms of the excited states of the non-interacting atomic QDOs, establishing a direct link between localized atomic fluctuations and delocalized collective modes. The full derivation of the MBD energy and ground-state correlation matrices is given in Ref.~\citenum{gori2023second}.

The key result for the present analysis is that this framework admits a canonical decomposition of the MBD energy into pairwise QDO contributions $E_{ij}$. From this decomposition, the contribution of atom $j$ to the force on atom $i$ along Cartesian direction $\alpha$ is
\begin{equation}
\label{eq:MBD_forces}
F_{i\alpha,\, j}
= -\frac{\partial E_{ij}}{\partial R_{i\alpha}},
\end{equation}
where $R_{i\alpha}$ is the $\alpha$-component of the position of atom $i$. By construction, $\sum_j F_{i\alpha,\,j} = F_{i\alpha}$ recovers the total MBD force on atom $i$, providing a physically transparent decomposition amenable to systematic analysis of interaction depth and directionality.

We note that several schemes have been developed to decompose the total electronic energy into atomic contributions~\cite{bader1990atoms, blanco2005interacting, bistoni2024local}. However, atomic energies are not quantum observables. In this work, instead, we focus on the decomposition of atomic forces. Atomic forces are rigorous quantum observables within the Born-Oppenheimer approximation, which provides a firm foundation for our analysis.

Quantum-mechanical Hilbert spaces of molecules admit a variety of states as a function of energy. These include bound, bound-in-continuum, and continuum states. Atom-centered basis sets used in electronic-structure calculations are approximate, in particular for the description of continuum states. Ultimately, studying strength and scatter of interatomic interactions should take into account all possible states in molecular Hilbert spaces. Here, we focus on (delocalized) bound states.

\subsection{Machine Learned and Empirical Force Fields}

To complement the SQ-MBD analysis with models of differing physical content, we trained two additional potentials. The first is PaiNN~\cite{schutt2021equivariant}, an equivariant message-passing neural network that can capture complex many-body effects without an explicit physical functional form. The second is a custom Mechanistic Empirical Force Field (MEFF) that adheres to the standard classical architecture: harmonic terms for covalent bonds and angles, and pairwise potentials for non-bonded interactions. Both models were optimized on MBD energy and force data using the SchNetPack package~\cite{schutt2019schnetpack, schutt2023schnetpack} for PaiNN MLFF and gradient descent for the MEFF.

Table~\ref{tab:model_accuracy} reports the accuracy of both models on held-out test sets. MLFF achieves sub-kcal/mol energy errors and force MAEs below 0.02~kcal/mol/\AA{} across all four systems, confirming that it faithfully learns the reference many-body interactions. The MEFF performs substantially worse, with errors growing systematically with system size (e.g., 3.91~kcal/mol energy MAE and 0.53~kcal/mol/\AA{} force MAE for Chignolin), reflecting the intrinsic limitations of fixed pairwise additive functional forms in describing non-additive many-body interactions.
 
To verify that the interaction patterns reported below are not specific to the long-range dispersion component, we separately trained MLFF on full \textit{ab initio} PBE0+MBD forces and repeated the pairwise decomposition. The resulting scatter and angular distributions are qualitatively similar to those obtained from the MBD-only decomposition (Fig.~\ref{fig:painn_full}, Tab.~\ref{tab:painn_full_table}), suggesting that long-range electron correlation is the dominant source of the interaction complexity we analyze. We therefore focus our analysis on the decomposition of MBD forces, for which the second-quantization framework provides an exact, physically grounded pairwise decomposition, while noting that the main conclusions hold for the full force landscape as well.

\begin{table}
\caption{Mean Absolute Errors (MAE) for energy and forces of the PaiNN MLFF and the Mechanistic Empirical Force Field (MEFF) trained on MBD reference data. Test sets include the tetrapeptide \ce{AcAla3NMe}, DHA, the Buckyball Catcher, and Chignolin. The MBD energy ranges are 6--44~kcal/mol and force components span $\pm$3.8~kcal/mol/\AA\ across the test sets; full dataset statistics are given in Tab.~\ref{tab:model_accuracy_mbd_extended}.}
\centering
\setlength{\tabcolsep}{4pt}
\begin{tabular}{l l cc}
\toprule
\textbf{System} & \textbf{Model} & \makecell{\textbf{Energy MAE}\\(kcal/mol)} & \makecell{\textbf{Force MAE}\\(kcal/mol/\AA)} \\
\midrule
\multirow{2}{*}{\ce{AcAla3NMe}} & MLFF & 0.04 & 0.008 \\
 & MEFF & 0.77 & 0.51 \\
\midrule
\multirow{2}{*}{DHA} & MLFF & 0.04 & 0.009 \\
 & MEFF & 1.62 & 0.70 \\
\midrule
\multirow{2}{*}{\makecell[l]{Buckyball\\Catcher}} & MLFF & 0.07 & 0.01 \\
 & MEFF & 3.30 & 0.84 \\
\midrule
\multirow{2}{*}{Chignolin} & MLFF & 0.13 & 0.02 \\
 & MEFF & 3.91 & 0.53 \\
\bottomrule
\end{tabular}
\label{tab:model_accuracy}
\end{table}

\begin{figure*}
    \centering
    \includegraphics[width=.99\linewidth]{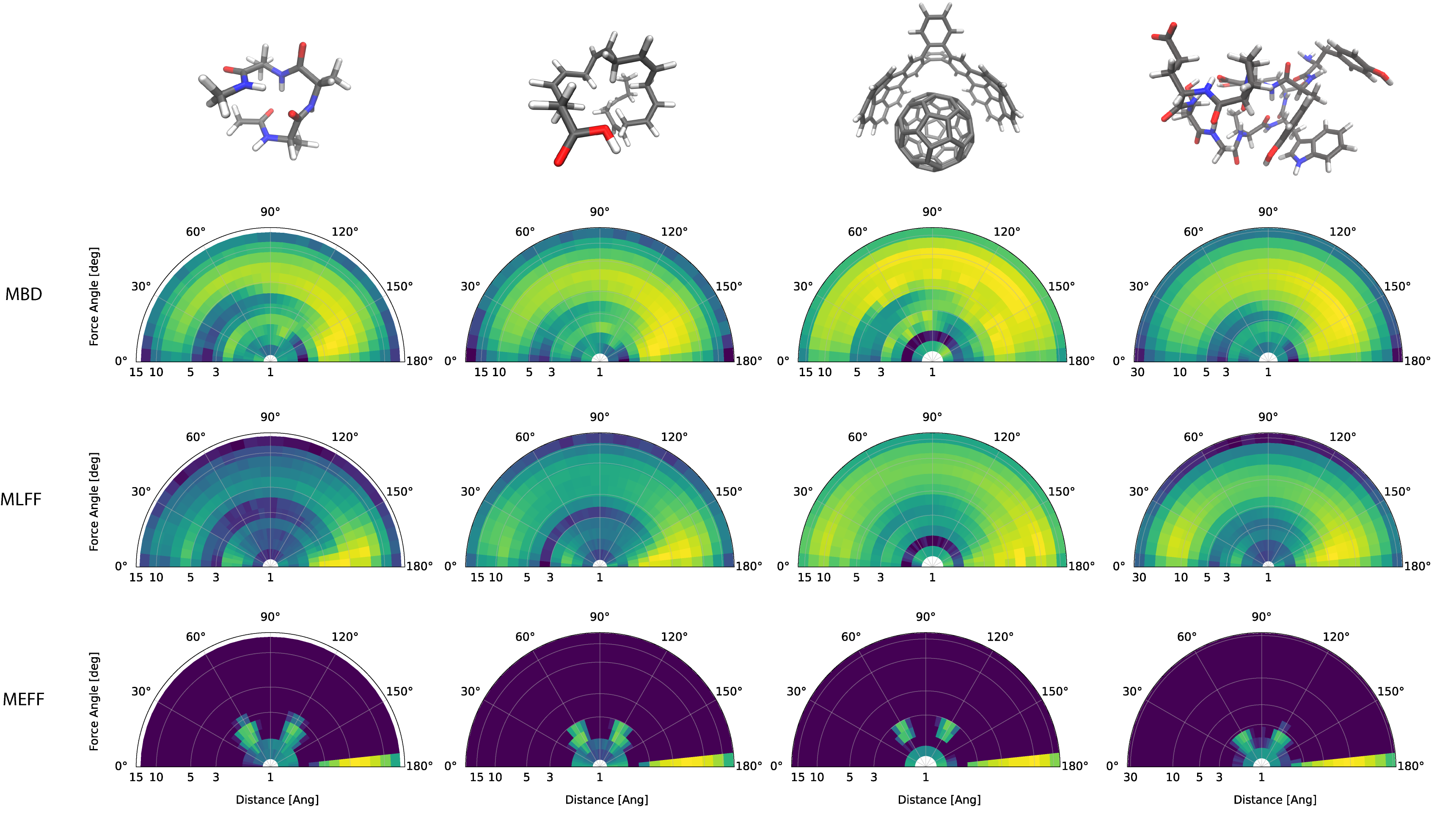}
    \caption{\textbf{Interaction anisotropy analysis.} Radial heatmaps displaying the angle between the pairwise force vector $F_{ij}$ and the interatomic distance vector ($180^\circ$ and $0^\circ$ represents perfect (anti)-alignment/isotropy). Columns compare results from SQ-MBD, MLFF, and MEFF. While the classical model shows rapid convergence to isotropy (uniform color), the ML and Quantum (MBD) models reveal rich angular structure and persistent anisotropy that increases with system size (rows). Colors encode the bin count on a logarithmic scale, with dark indicating low counts and light indicating high counts.}
    \label{fig:radial_heatmaps}
\end{figure*}

\subsection{Interaction Anisotropy}

We begin by quantifying the directional character of pairwise forces via the angle $\theta_{ij}$ between the force contribution vector $F_{ij}$ and the interatomic displacement vector $d_{ij}$ (Fig.~\ref{fig:overview}A, C). In a purely isotropic, pairwise-additive framework, all force contributions would point along $d_{ij}$ ($\theta_{ij} = 180^\circ$ for attractions), yielding a delta-function angular distribution at all distances. Any such deviation, expressed as a spread in $\theta_{ij}$ values, is a direct signature of the non-local electronic correlations encoded in the quantum state.

The angular distributions derived from SQ-MBD and MLFF (Fig.~\ref{fig:radial_heatmaps}) show a clear trend: anisotropy not only persists at long-range but grows with molecular size. For the 42-atom tetrapeptide \ce{AcAla3NMe}, significant angular spread is already visible, challenging the assumption that anisotropic effects are negligible in simple biomolecules. For DHA (56 atoms) and the Buckyball Catcher (148 atoms), this spread amplifies substantially and extends to larger interatomic separations. The angular distribution of the Chignolin protein (166 atoms) exhibits the broadest and most distance-persistent anisotropy in our test set. This trend implies that as molecular complexity grows, delocalized electronic fluctuations couple across the supramolecular scaffold, generating force components transverse to the interatomic axis that effectively ``steer'' atoms in directions that are not captured by isotropic models.

The symmetry of the Buckyball Catcher complex, which comprises two corannulene-based buckybowls enclosing \ce{C60}, offers an instructive special case - the angular distributions reflect the high local symmetry of the host--guest interaction geometry, with anisotropy concentrated along symmetry-inequivalent interaction channels. This provides further evidence that the observed anisotropy encodes the global structure of the molecule.

The MLFF qualitatively reproduces these angular features (Fig.~\ref{fig:radial_heatmaps}, MLFF row), captures some degree of directional structure and correctly identifies systems with higher anisotropy. However, the angular distribution predicted by MLFF is noticeably more limited than the SQ-MBD reference, particularly at longer distances. This limitation is again attributable to the semi-local MPNN architecture: subtle, long-range polarization effects that propagate across the full molecular graph cannot be faithfully represented when information flow is restricted to a finite receptive field. The MEFF, as expected, shows rapid convergence to isotropy with increasing distance, with the angular distribution effectively collapsing to $\theta_{ij} = 180^\circ$ beyond a few \AA ngstr\"{o}ms (Fig.~\ref{fig:radial_heatmaps}, MEFF row), consistent with its pairwise additive functional form.

\begin{figure*}
    \centering
    \includegraphics[width=.99\linewidth]{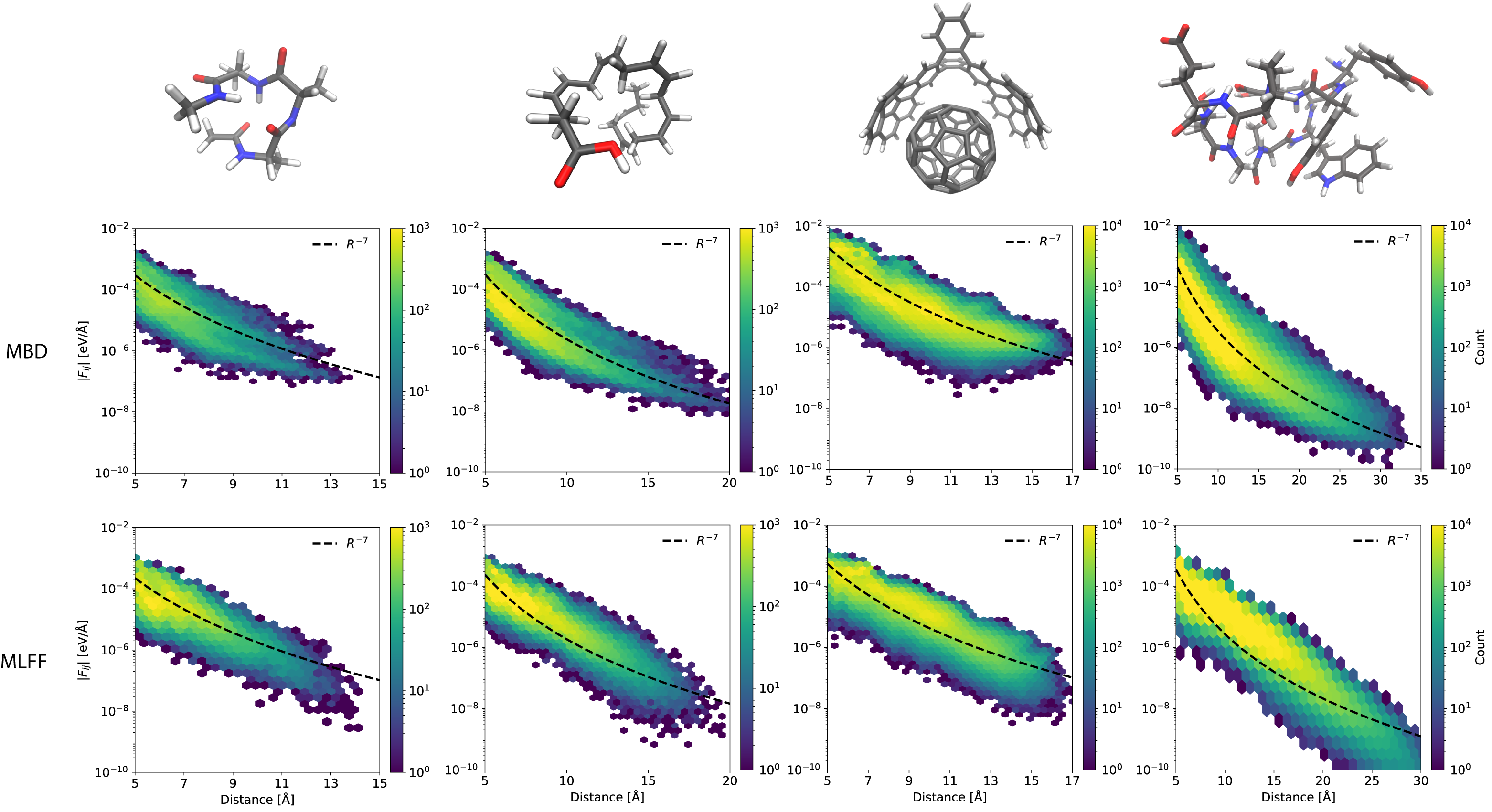}
    \caption{\textbf{Interaction strength decay.} Dispersion force contributions vs. interatomic distance for SQ-MBD and MLFF. The dashed reference curve is a fit of $|F_{ij}| = c\,R^{-7}$ over the $5$--$10$~\text{Å} range.
    } 
    \label{fig:magnitude}
\end{figure*}

\subsection{Interaction Depth}

Complementing the interaction anisotropy analysis, we now examine the interaction depth, defined as the magnitude of the pairwise force contribution $|F_{ij}|$, as a function of the interatomic distance in the presence of all other atoms. Conventional dispersion force fields assume $|F_{ij}| \propto R^{-7}$, the derivative of London $R^{-6}$ dispersion energy, which would collapse all interactions into a single monotonic curve for each pair of elements. Instead, the raw interaction data from SQ-MBD reveal a broad ``cloud'' of interaction strengths spanning several orders of magnitude at any given distance (Fig.~\ref{fig:magnitude}, MBD row). Notably, the distribution exhibits significant positive deviations at long-range: specific atom pairs separated by distances as large as 20~\text{\AA} can experience forces comparable to, or even exceeding, those between pairs at 10~\text{\AA}. 

The broad scatter in interaction magnitudes is likewise qualitatively reproduced by MLFF (Fig.~\ref{fig:magnitude}, MLFF row). Despite lacking an explicit physics-based dispersion functional form, MLFF learns to predict the same enhanced scatter observed in the MBD reference data, confirming that this behavior is a learnable feature of the potential energy landscape. However, as a semi-local message-passing network, MLFF's ability to capture long-range scatter is inherently limited by its receptive field. In our setup, the cutoff and number of layers were chosen to connect the most distant atom pairs in the training structures, whereas conventional MLFFs typically use 2--3 message-passing layers at a 5~\text{\AA} cutoff. Beyond approximately 10--15~\text{\AA}, atom pairs lie entirely outside such a network's receptive field and their predicted interactions undergo an effective exponential decay~\cite{esders2025analyzing}. That MLFF nonetheless captures the correct scatter behavior within its receptive field partly rationalizes the empirical success of MLFFs on complex molecular systems while also pinpointing a clear limitation for capturing truly long-range interactions.

 \begin{figure*}
     \centering
     \includegraphics[width=1.0\linewidth]{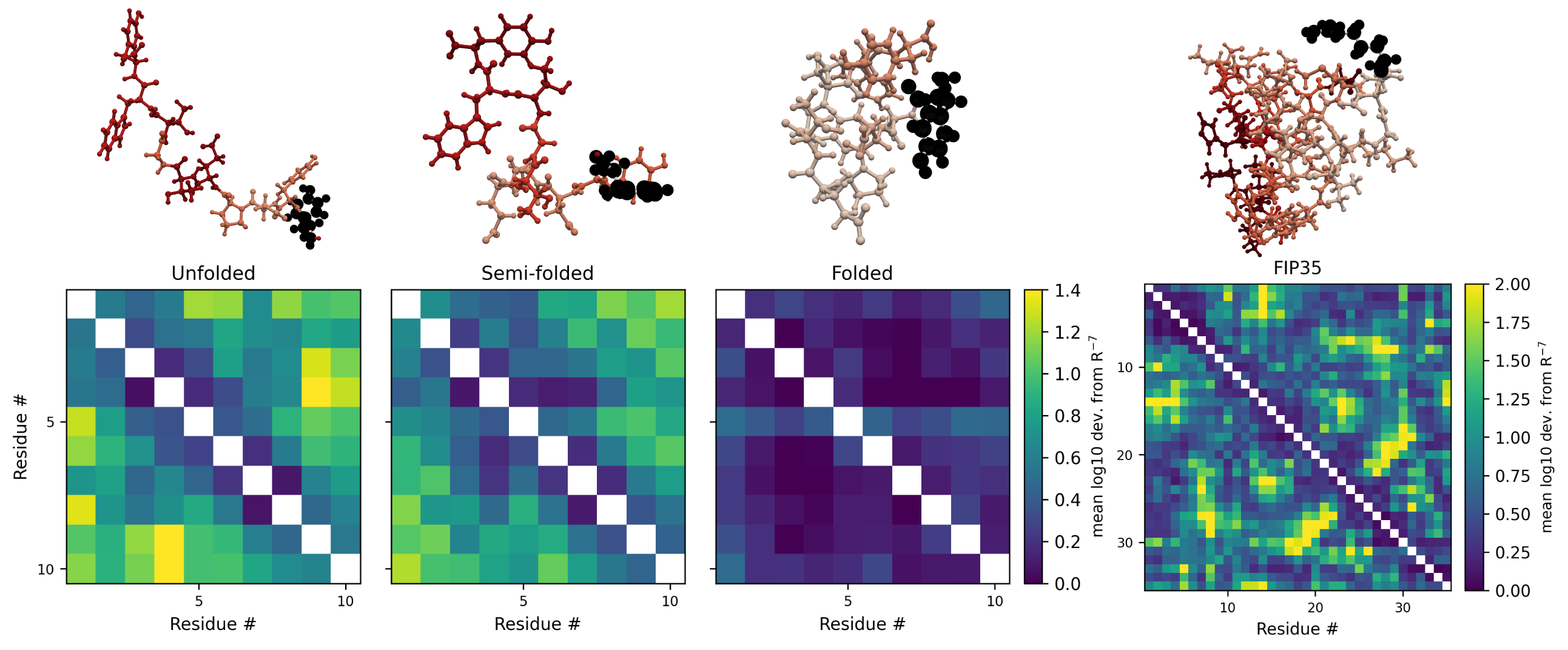}
     \caption{\textbf{Interaction hotspots.} Interaction analysis of Chignolin across folding states (unfolded, semi-folded, and folded; 10 residues) and the FIP35 protein (35 residues). Top row: molecular visualizations colored by the force-weighted average $\log_{10}$ deviation from the $R^{-7}$ power law (red encodes strong deviations and white encodes pairwise-like behavior); TYR1 for Chignolin and ARG14 for FIP35 are highlighted in black. Bottom row: per-residue mean $\log_{10}$ deviation from $R^{-7}$ (zero corresponds to purely pairwise behavior, positive values indicate excess non-pairwise interaction). Per-residue deviations span $[-0.21, 1.60]$ across the Chignolin panels and $[0.02, 3.18]$ for FIP35; the colormap is capped at $1.4$ for Chignolin and $2.0$ for FIP35. See Methods for details on the per-residue interaction decomposition.}
     \label{fig:chig_fragment}
 \end{figure*}

These observations suggest that in proteins and similarly complex systems, long-range dispersion forces remain collectively relevant well beyond typical cutoff distances and may deviate considerably from the $R^{-7}$ pairwise law. 
The latter has also been reported for protein-nanowire complexes, where the many-body interaction energy decays considerably more slowly than pairwise-additive models predict, with the deviation growing with separation~\cite{ambrosetti2016wavelike}.

\subsection{Interaction Hotspots}

To visualize the structural origin of the scatter, we project the $R^{-7}$ deviations onto individual atoms and residues (Fig.~\ref{fig:chig_fragment}). For Chignolin, we trace the interaction landscape across three representative folding states (unfolded, semi-folded, and folded)~\cite{sali1994how}, each analyzed over $\sim$33 conformations drawn from the 100-snapshot trajectory and partitioned by principal moment of inertia.

The per-residue interaction heatmaps reveal that the distribution of deviations from $R^{-7}$ scaling is far from uniform and changes markedly with the folding state. In the unfolded configuration, where the chain is extended and interatomic separations are large, the heatmap exhibits the most pronounced deviations, with strong off-diagonal features indicating that distant residue pairs interact up to 40 times stronger than a pairwise power law would predict ($10^{1.6}$). As the protein progresses through the semi-folded state toward the fully folded $\beta$-hairpin, the deviations systematically diminish and the interaction decay converges closer to the default $R^{-7}$ scaling.

To test whether these features persist in larger proteins, we extend the analysis to FIP35 (562 atoms, 35 residues). The per-residue heatmap for FIP35 (Fig.~\ref{fig:chig_fragment}, last column) displays qualitatively similar behavior: the deviations from $R^{-7}$ are strongly heterogeneous across residue pairs, with distinct hotspot clusters appearing at long intramolecular separations. Notably, the deviations in FIP35 are larger than in Chignolin, with the mean log$_{10}$ deviation reaching a maximum of $\sim$3.1 for SER2--ARG14 (i.e. more than 1000 times stronger than the expected pairwise force) compared to $\sim$1.6 for  PRO4--TRP9 in Chignolin. This is consistent with the expectation that the many-body character of dispersion interactions amplifies with system size and structural complexity. The FIP35 heatmap also reveals a richer off-diagonal structure, reflecting the more complex tertiary fold and the greater diversity of long-range residue-residue contacts available in a 35-residue protein. These results suggest that the interaction hotspot phenomenon is not specific to the minimal Chignolin system but is a general feature of protein-sized molecules, and that its magnitude grows with molecular size.

\section{Discussion}

Our analysis demonstrates that long-range interatomic forces within molecules cannot be understood as smooth, isotropic, pairwise-additive functions of distance, but rather emerge from the collective quantum-mechanical response of the entire electronic system. Across five systems spanning an order of magnitude in size, we find that: (i) the angular distribution of pairwise forces remains broad well beyond conventional cutoffs, (ii) the interaction depth at fixed distance scatters by up to four orders of magnitude, and (iii) both scatter and anisotropy are amplified with system size. 

These features have direct consequences for systems in which long-range coupling between distant fragments is expected to matter, from the folding pathways of miniproteins such as Chignolin, to the conformational plasticity of larger domains such as FIP35, to host-guest recognition in supramolecular complexes exemplified by the Buckyball Catcher. Our results motivate a shift in the objects of molecular modeling, from interacting atoms to interacting \emph{hotspots}: extended regions whose electronic response is collectively determined and whose mutual coupling is anisotropic and distance-persistent. Rather than spreading uniformly across a protein, hotspots concentrate on specific residues and residue pairs, and the pattern changes markedly with conformational state. Collective electronic fluctuations therefore provide a long-range, directional coupling between distant fragments, with potential implications for folding kinetics~\cite{lindorff2011fast}, ligand recognition~\cite{hermann2017first}, and allosteric regulation~\cite{motlagh2014ensemble}.

These observations also have direct consequences for force-field design. Mechanistic force fields with fixed pairwise functional forms collapse to the isotropic limit within a few angstroms. Machine learning force fields, by contrast, recover much of the correct scatter and anisotropy within their receptive field, which rationalizes their success in simulating complex molecular behavior. Their semi-local architecture, however, causes them to either ignore long-range interactions beyond the receptive field or revert to a mean-field description through pairwise electrostatics and dispersion modules. Closing this gap defines a roadmap for quantum-aware force fields, to be built either from architectures that represent delocalized collective modes natively, or from hybrid schemes in which local learned components are combined with explicit non-local physics.

\section{Methods}
\textbf{Systems}.
Molecular geometries for \ce{AcAla3NMe}, DHA, and the Buckyball Catcher were taken from the MD22 dataset~\cite{chmiela2023accurate}; Chignolin geometries were taken from the AIMD-Chig dataset~\cite{wang2023aimd}; FIP35 geometries were taken from Refs.~\citenum{stohr2019quantum, lindorff2011fast}.

\textbf{Reference calculations}. 
The DFT+MBD and MBD energies and forces used for training MLFF and MEFF were taken from the MD22 dataset for \ce{AcAla3NMe}, DHA, and the Buckyball Catcher. 

For Chignolin, 462 frames were calculated at the PBE0+MBD level of theory using the FHI-aims code~\cite{blum2009ab, ren2012resolution}, with ``tight'' settings for basis functions and integration grids. Energies were converged to $10^{-5}$~eV, force accuracy to $10^{-4}$~eV/\AA, and the SCF convergence criteria were $10^{-5}$~eV for the sum of eigenvalues and $10^{-3}$~electrons/\AA$^3$ for the charge density. 

\textbf{Hirshfeld ratios for MBD@rsSCS.} 
The range-separated self-consistent screening MBD method (MBD@rsSCS)~\cite{tkatchenko2012accurate, distasio2014many} requires per-atom Hirshfeld volume ratios as input, 
\begin{equation*} 
h_i = \frac{V_i^{\mathrm{AIM}}}{V_i^{\mathrm{free}}}, 
\end{equation*} 
where $V_i^{\mathrm{AIM}}$ is the Hirshfeld-partitioned atomic volume of atom $i$ in the molecule and $V_i^{\mathrm{free}}$ is the corresponding free-atom reference. These ratios rescale the free-atom static polarizabilities and $C_6$ coefficients via $\alpha_i = h_i\,\alpha_i^{\mathrm{free}}$ and $C_{6,ii} = h_i^{2}\,C_{6,ii}^{\mathrm{free}}$, providing the parameters of the QDO Hamiltonian in Eq.~\eqref{eq:SI_MBD_H}. Because the SQ-MBD pairwise force decomposition relies on a finite-difference scheme requiring $6N+1$ evaluations of $\{h_i\}$ per frame (see Eq.~\eqref{eq:SI_MBD_forces}), Hirshfeld ratios were re-evaluated at every displaced geometry.

For the Buckyball Catcher (148 atoms, 123 conformations), $h_i$ values were obtained from PBE+MBD calculations using the FHI-aims code~\cite{blum2009ab, ren2012resolution}, with ``light'' settings for basis functions and integration grids. Energies were converged to $10^{-6}$~eV, force accuracy to $5\times10^{-4}$~eV/\AA, and the SCF convergence criteria were $10^{-5}$~eV for the sum of eigenvalues and $10^{-4}$~electrons/\AA$^3$ for the charge density. This corresponds to $6N+1 = 889$ evaluations per frame, totaling $\sim\!10^{5}$ single-point calculations over the $\sim$123-frame ensemble.

For the (bio)molecular systems (\ce{AcAla3NMe}, 42 atoms, 86 conformations; DHA, 56 atoms, 61 conformations; Chignolin, 166 atoms, 100 conformations; FIP35, 562 atoms, 3 conformations), $h_i$ values were predicted by the SO3LR MLFF with float64 precision~\cite{kabylda2025molecular}, which was trained on a diverse dataset of organic systems and reproduces \emph{ab initio} Hirshfeld ratios with high fidelity. The use of SO3LR was essential here, as the corresponding $6N+1$ factors ($\sim$1000 for Chignolin, $\sim$3400 for FIP35) would have rendered explicit \emph{ab initio} ratio evaluation prohibitive.

Furthermore, because long-range pair interactions are subtle and small in magnitude, the underlying DFT calculations must be tightly converged for the resulting ratios to yield signal above the intrinsic numerical noise floor. This requirement becomes very challenging, and in practice often infeasible, for systems beyond $\sim$150 atoms.

\textbf{Second quantization of many-body dispersion: full derivation.}
A primary limitation of density functional theory (DFT) is its difficulty in capturing atomic forces arising from long-range correlations in the electron density. These correlations originate from Coulomb interactions and are typically missing from conventional density-functional approximations. The many-body dispersion (MBD) formalism~\cite{tkatchenko2012accurate, distasio2014many} provides an accurate and efficient estimate of these correlation-energy contributions, identifiable as many-body van der Waals interactions. In the MBD model, each atom $i$ is represented by a quantum harmonic oscillator carrying an electric dipole, a Quantum Drude Oscillator (QDO), parameterized to reproduce the electric-response properties of atoms-in-molecules. The atomic QDOs interact through dipole--dipole couplings, leading to the Hamiltonian
\begin{equation}
\label{eq:SI_MBD_H}
\hat{H}_{\mathrm{MBD},\lambda}
=\frac{1}{2}\sum_{i=1}^N
\left[
\|\hat{\boldsymbol{p}}_i\|^2
+\frac{1}{2}\sum_{j}
\hat{\boldsymbol{q}}_i\,\mathbf{V}_{ij}\,\hat{\boldsymbol{q}}_j
\right],
\end{equation}
where $\hat{\boldsymbol{q}}_i=\sqrt{m_i}\,\hat{\boldsymbol{r}}_i$ is the mass-weighted displacement operator of atom $i$ and $\hat{\boldsymbol{p}}_i$ its conjugate momentum. The potential interaction matrix is $\mathbf{V}_{ij}= \omega_i^2\, \mathbf{I}_{3\times3}\,\delta_{ij}+\lambda\, \mathbf{T}_{ij}(1-\delta_{ij})$, with $\mathbf{I}_{3\times3}$ the identity matrix and $\mathbf{T}_{ij}(\boldsymbol{R}_{ij})$ the dipole--dipole interaction tensor. The coupling parameter $\lambda \in [0,1]$ interpolates between the non-interacting ($\lambda=0$) and fully interacting ($\lambda=1$) systems.

The MBD interaction energy is defined as the difference between the ground-state energies of the interacting and non-interacting QDO systems,
\begin{equation}
\label{eq:SI_MBD_energy}
E_{\mathrm{MBD}}
=
\langle\Psi_{\mathrm{GS},1}|\hat{H}_{\mathrm{MBD},1}|\Psi_{\mathrm{GS},1}\rangle
-\langle\Psi_{\mathrm{GS},0}|\hat{H}_{\mathrm{MBD},0}|\Psi_{\mathrm{GS},0}\rangle.
\end{equation}
In general, $E_{\mathrm{MBD}}$ is a non-additive function of the atomic positions. Its many-body character is encoded in the ground-state correlation matrices of QDO displacements and momenta,
\begin{align}
\Gamma^{(qq)}_{ij,\lambda}
&=\langle\Psi_{\mathrm{GS},\lambda}|\hat{\boldsymbol{q}}_i\!\otimes\!\hat{\boldsymbol{q}}_j|\Psi_{\mathrm{GS},\lambda}\rangle,\\
\Gamma^{(pp)}_{ij,\lambda}
&=\langle\Psi_{\mathrm{GS},\lambda}|\hat{\boldsymbol{p}}_i\!\otimes\!\hat{\boldsymbol{p}}_j|\Psi_{\mathrm{GS},\lambda}\rangle.
\end{align}
The entries of the $3\times3$ blocks of $\Gamma^{(qq)}_{ij,\lambda}$ and $\Gamma^{(pp)}_{ij,\lambda}$ depend, in general, on the coordinates of all atomic centers, reflecting the genuinely non-local nature of the interaction.

Expressing the MBD energy in terms of these correlation matrices yields a canonical decomposition of $E_{\rm MBD}$ into single-atom and pairwise QDO contributions:
\begin{align}
E_{ii} &= \frac{1}{2}\sum_{\alpha=1}^3 \left[(\Gamma^{(pp)}_{ii,1})_{\alpha\alpha}
        + \omega_i^2 (\Gamma^{(qq)}_{ii,1})_{\alpha\alpha} - \hbar \omega_i \right],\\
E_{ij} &= \frac{1}{2}\sum_{\alpha,\beta=1}^3 (\Gamma^{(qq)}_{ij,1})_{\alpha\beta}\,
        T_{ij,\alpha\beta}.
\end{align}
Expectation values of polynomial observables in the MBD ground state, including these correlation matrices, are efficiently computed within the second-quantization framework introduced in Ref.~\citenum{gori2023second}, which is inspired by quantum field theoretical methods and makes explicit the linear (Bogoliubov) transformation between the atomic QDO operators and the collective MBD modes.

The pairwise force contribution is obtained by differentiating the decomposed energy with respect to atomic coordinates. In practice, these derivatives are evaluated numerically via central finite differences:
\begin{equation}
\label{eq:SI_MBD_forces}
F_{ij,\alpha}
= -\frac{\partial E_{ij}}{\partial R_{i,\alpha}}
\approx -\frac{E_{ij}(\boldsymbol{R}+\Delta\,\boldsymbol{e}_{i,\alpha})
              - E_{ij}(\boldsymbol{R}-\Delta\,\boldsymbol{e}_{i,\alpha})}{2\Delta},
\end{equation}
where $\Delta = 0.001$~\AA{} is a small displacement of atom $i$ along the Cartesian unit vector $\boldsymbol{e}_{i,\alpha}$, and $E_{ij}$ is evaluated at each perturbed geometry using the framework of Ref.~\citenum{gori2023second}. By construction, $\sum_j F_{ij,\alpha} = F_{i,\alpha}$ recovers the total MBD force on atom $i$, providing a physically transparent decomposition amenable to systematic analysis of interaction depth and directionality.

\textbf{PaiNN MLFF training details.} 
All PaiNN models were trained using the SchNetPack package~\cite{schutt2019schnetpack, schutt2023schnetpack} with standard parameters as reported in the original publication~\cite{schutt2021equivariant}. The embedding size was $128$, and $30$ Gaussian radial basis functions were used to embed pairwise distances. The networks used $3$ interaction layers, with cutoffs $8$~\AA{} for \ce{AcAla3NMe}, $10$~\AA{} for DHA, $8$~\AA{} for the Buckyball Catcher, and $12$~\AA{} for Chignolin. Optimization used Adam with a learning rate of $0.0005$ and weight decay of $0.001$.  Data split sizes were $n_{\text{train}}=950$, $n_{\text{val}}=50$ and the rest of each dataset used for testing.

\textbf{PaiNN Force Decomposition.} To predict molecular energies, message-passing neural networks such as PaiNN decompose the total energy into per-atom contributions. For a molecule of $N$ atoms, each atom $i$ is assigned a scalar energy contribution $E_i$, and the total energy is recovered as their sum,
\begin{equation}
E = \sum_{i=1}^{N} E_i.
\end{equation}

The total force on atom $i$ is the negative gradient of the total energy with respect to its position $\mathbf{R}_i$. This is not only a physical requirement but also the operational definition used by the network, which obtains forces by automatic differentiation of the predicted energy:
\begin{equation}
\mathbf{F}_i
= -\frac{\partial E}{\partial \mathbf{R}_i}
= -\sum_{j=1}^{N} \frac{\partial E_j}{\partial \mathbf{R}_i}.
\end{equation}

This expression admits a natural pairwise decomposition. We define the contribution of atom $j$ to the force on atom $i$ as
\begin{equation} 
\mathbf{F}_{ij} = -\frac{\partial E_j}{\partial \mathbf{R}_i}, 
\end{equation}
so that $\sum_{j} \mathbf{F}_{ij} = \mathbf{F}_i$ by construction, in direct analogy with the SQ-MBD decomposition of Eq.~\eqref{eq:MBD_forces}. The interaction strength between atoms $i$ and $j$ is then defined as the Euclidean norm of this vector,
\begin{equation}
F_{ij} = \|\mathbf{F}_{ij}\|_2
       = \left(\sum_{\alpha=1}^{3} F_{ij,\alpha}^{2}\right)^{\!1/2},
\end{equation}
where $F_{ij,\alpha}$ denotes the Cartesian component of $\mathbf{F}_{ij}$ along direction $\alpha$.

\textbf{Empirical Force Field Decomposition.}
Empirical force fields predict forces via a many-body decomposition of the total energy. This decomposition typically includes terms up to fourth order:
\begin{equation}
\begin{aligned}
E_{\text{tot}} = &\sum_i E^{(1)}(i)
+ \sum_{i<j} E^{(2)}(i,j) \\
&+ \sum_{i<j<k} E^{(3)}(i,j,k)
+ \sum_{i<j<k<l} E^{(4)}(i,j,k,l),
\end{aligned}
\end{equation}
where $E^{(1)}$ is the one-body energy term, $E^{(2)}$ the two-body term, and so on. The functional form chosen for this study is
\begin{equation}
\begin{aligned}
E_{\text{tot}} =\;
&\sum_{\substack{i,j \\ \text{bonded}}}
   k_{R_{ij}}\bigl(R_{ij} - R_{0,ij}\bigr)^2 \\
&+ \sum_{\substack{i,j,k \\ \text{bonded}}}
   k_{\theta_{ijk}}\bigl(\theta_{ijk} - \theta_{0,ijk}\bigr)^2 \\
&- \sum_{\substack{i,j \\ \text{not 1,3-bonded}}}
   \frac{k_{C}\bigl(\mathrm{elem}(i),\mathrm{elem}(j)\bigr)}{R_{ij}^{6}} \\
&+ \sum_{\substack{i,j \\ \text{not 1,3-bonded}}}
   \frac{k_q\, q_i q_j}{R_{ij}},
\end{aligned}
\end{equation}
where $R_{ij}$ is the distance between atoms $i$ and $j$; $k_{R_{ij}}$ is a trainable force constant specific to the $(i,j)$ pair; $\theta_{ijk}$ is the angle at atom $j$ in triplets where $i$ is bonded to $j$ and $j$ is bonded to $k$; $k_{\theta_{ijk}}$ is the trainable force constant for the $(i,j,k)$ triplet; $k_C$ is the trainable dispersion coefficient specific to the elements of $i$ and $j$; and $k_q$ is the trainable Coulomb coefficient, made trainable to account for the effective permittivity of the various tested molecules. In the experiments where the empirical force field is fitted to MBD-only data, the Coulomb term is excluded from the functional form. The force field is trained by gradient descent on the same data as PaiNN, using a learning rate of $0.002$, until convergence.

To obtain a pairwise decomposition consistent with the PaiNN scheme above, we partition the total energy into per-atom contributions $E_i$ such that $\sum_i E_i = E_{\text{tot}}$, by distributing each $n$-body term equally among the $n$ atoms involved:
\begin{equation}
\begin{aligned}
E_i = \;& E^{(1)}(i) + \tfrac{1}{2} \sum_{j\neq i} E^{(2)}(i,j) \\
        &+ \tfrac{1}{3}\!\!\sum_{\substack{j<k \\ j,k \neq i}}\!\! E^{(3)}(i,j,k)
         + \tfrac{1}{4}\!\!\sum_{\substack{j<k<l \\ j,k,l \neq i}}\!\! E^{(4)}(i,j,k,l).
\end{aligned}
\end{equation}
The pairwise force contribution from atom $j$ to the force on atom $i$ is then defined in direct analogy with the PaiNN case as
\begin{equation}
\mathbf{F}_{ij} = -\frac{\partial E_j}{\partial \mathbf{R}_i},
\end{equation}
with corresponding interaction strength $F_{ij} = \|\mathbf{F}_{ij}\|_2$.

\textbf{Per-residue interaction decomposition.}
Interaction analysis was performed for Chignolin across folding states and for FIP35. Molecular structures in Fig.~\ref{fig:chig_fragment} are colored by the force-weighted, frame-averaged per-atom deviation from a pairwise $R^{-7}$ baseline,
\begin{equation*}
\Delta_{ij}^{(\mathrm{atom})}
 = \log\!\left(
        \frac{\langle F_{ij}\rangle\,\langle R_{ij}\rangle^{7}}
             {C_{6}^{(z_i,z_j)}}
   \right),
\end{equation*}
where $\langle\cdot\rangle$ denotes the average over frames, $z_i$ is the element of atom $i$, and $C_{6}^{(z_i,z_j)}$ is the element-pair dispersion coefficient fitted from all atom pairs of that type at $R_{ij} \geq 5$~\AA{} (so $\Delta = 0$ for purely pairwise behavior and $\Delta > 0$ for excess non-pairwise interaction). Red encodes strong deviations, white encodes pairwise-like behavior; the probe residue (TYR1 for Chignolin, ARG14 for FIP35) is highlighted in black. The per-residue-pair heatmaps show the mean of $\Delta_{ij}^{(\mathrm{atom})}$ over all atom pairs $(i,j)$ bridging residues $a$ and $b$,
\begin{equation*}
\Delta_{ab}^{(\mathrm{res})}
 = \frac{1}{N_{ab}}\sum_{i\in a,\,j\in b}\Delta_{ij}^{(\mathrm{atom})},
\end{equation*}
with the diagonal ($a=b$) masked.

\section*{Funding}
A.K. acknowledges financial support from the Luxembourg National Research Fund (FNR AFR Ph.D. Grant 15720828). 
K.R.M. was supported in part by the German Ministry for Education and Research (BMBF) under Grants 01IS14013A-E, 01GQ1115, 01GQ0850, 01IS18025A, 031L0207D, 01IS18037A; by the Institute of Information \& Communications Technology Planning \& Evaluation (IITP) grant funded by the Korea government (MSIT) (No. RS-2019-II190079, Artificial Intelligence Graduate School Program, Korea University); by the Korea government (MSIT) (No. RS-2024-00457882, AI Research Hub Project). A.T. and M.G. acknowledges the Luxembourg National Research Fund under grant FNR-CORE MBD-in-BMD and the European Research Council under ERC-AdG grant FITMOL.

\section*{Author contributions}
A.K. and A.T. conceived and designed the study. M.G. developed and implemented the SQ-MBD formalism. A.K. extended the SQ-MBD formalism to forces and performed the reference calculations. M.E. trained the MLFF and MEFF models and performed the corresponding force decompositions. A.K. analyzed the interaction depth, anisotropy, and hotspot patterns and produced the figures. S.C. contributed to the interpretation of the results. A.K. and A.T. wrote the manuscript with input from all authors. A.T. and K.R.M. supervised the project and acquired funding. All authors discussed the results and contributed to editing the manuscript. Correspondence should be
addressed to A.T. and K.R.M.

\section*{Competing interests}
The authors declare no competing interests.

\bibliography{references}

@article{gurwitsch1914physiko,
  title={{\"U}ber die physiko-chemische Attraktionskraft},
  author={Gurwitsch, L},
  journal={Z. physik. Chem.},
  volume={87},
  pages={323--332},
  year={1914},
  publisher={De Gruyter Oldenbourg},
  doi={10.1515/zpch-1914-8720}
}

@article{langmuir1917constitution,
  title={The constitution and fundamental properties of solids and liquids. II. Liquids.},
  author={Langmuir, Irving},
  journal={J. Am. Chem. Soc.},
  volume={39},
  number={9},
  pages={1848--1906},
  year={1917},
  publisher={ACS Publications},
  doi={10.1021/ja02254a006}
}

@article{price2000toward,
  title={Toward more accurate model intermolecular potentials for organic molecules},
  author={Price, Sarah L},
  journal={Rev. Comput. Chem.},
  volume={14},
  pages={225--289},
  year={2000},
  doi={10.1002/9780470125915.ch4}
}

@article{stone1988some,
  title={Some new ideas in the theory of intermolecular forces: anisotropic atom-atom potentials},
  author={Stone, AJ and Price, SL},
  journal={J. Phys. Chem.},
  volume={92},
  number={12},
  pages={3325--3335},
  year={1988},
  publisher={ACS Publications},
  doi={10.1021/j100323a006}
}

@article{kriz2024quantification,
  title={Quantification of Anisotropy in Exchange and Dispersion Interactions: A Simple Model for Physics-Based Force Fields},
  author={Kriz, Kristian and Van der Spoel, David},
  journal={J. Phys. Chem. Lett.},
  volume={15},
  number={39},
  pages={9974--9978},
  year={2024},
  publisher={ACS Publications},
  doi={10.1021/acs.jpclett.4c02034}
}

@article{dinur1989direct,
  title={Direct evaluation of nonbonding interactions from ab initio calculations},
  author={Dinur, Uri and Hagler, Arnold T},
  journal={J. Am. Chem. Soc.},
  volume={111},
  number={14},
  pages={5149--5151},
  year={1989},
  publisher={ACS Publications},
  doi={10.1021/ja00196a021}
}

@article{van2018new,
  title={New angles on standard force fields: Toward a general approach for treating atomic-level anisotropy},
  author={Van Vleet, Mary J and Misquitta, Alston J and Schmidt, JR},
  journal={J. Chem. Theory Comput.},
  volume={14},
  number={2},
  pages={739--758},
  year={2018},
  publisher={ACS Publications},
  doi={10.1021/acs.jctc.7b00851}
}

@article{tkatchenko2012accurate,
  title={Accurate and efficient method for many-body van der Waals interactions},
  author={Tkatchenko, Alexandre and DiStasio Jr, Robert A and Car, Roberto and Scheffler, Matthias},
  journal={Phys. Rev. Lett.},
  volume={108},
  number={23},
  pages={236402},
  year={2012},
  publisher={APS},
  doi={10.1103/PhysRevLett.108.236402}
}

@article{distasio2014many,
  title={Many-body van der Waals interactions in molecules and condensed matter},
  author={DiStasio, Robert A and Gobre, Vivekanand V and Tkatchenko, Alexandre},
  journal={J. Phys.: Condens. Matter},
  volume={26},
  number={21},
  pages={213202},
  year={2014},
  publisher={IOP Publishing},
  doi={10.1088/0953-8984/26/21/213202}
}

@article{gori2023second,
  title={Second quantization of many-body dispersion interactions for chemical and biological systems},
  author={Gori, Matteo and Kurian, Philip and Tkatchenko, Alexandre},
  journal={Nat. Commun.},
  volume={14},
  number={1},
  pages={8218},
  year={2023},
  publisher={Nature Publishing Group UK London},
  doi={10.1038/s41467-023-43785-z}
}

@article{esders2025analyzing,
  title={Analyzing atomic interactions in molecules as learned by neural networks},
  author={Esders, Malte and Schnake, Thomas and Lederer, Jonas and Kabylda, Adil and Montavon, Gr{\'e}goire and Tkatchenko, Alexandre and M{\"u}ller, Klaus-Robert},
  journal={J. Chem. Theory Comput.},
  volume={21},
  number={2},
  pages={714--729},
  year={2025},
  publisher={ACS Publications},
  doi={10.1021/acs.jctc.4c01424}
}

@article{unke2021machine,
  title={Machine learning force fields},
  author={Unke, Oliver T and Chmiela, Stefan and Sauceda, Huziel E and Gastegger, Michael and Poltavsky, Igor and Sch{\"u}tt, Kristof T and Tkatchenko, Alexandre and M{\"u}ller, Klaus-Robert},
  journal={Chem. Rev.},
  volume={121},
  number={16},
  pages={10142--10186},
  year={2021},
  publisher={ACS Publications},
  doi = {10.1021/acs.chemrev.0c01111}
}

@article{schutt2023schnetpack,
    author = {Sch{\"u}tt, Kristof T. and Hessmann, Stefaan S. P. and Gebauer, Niklas W. A. and Lederer, Jonas and Gastegger, Michael},
    title = "{SchNetPack 2.0: A neural network toolbox for atomistic machine learning}",
    journal = {J. Chem. Phys.},
    volume = {158},
    number = {14},
    pages = {144801},
    year = {2023},
    month = {04},
    doi = {10.1063/5.0138367},
}

@article{schutt2019schnetpack,
    author = {Sch{\"u}tt, Kristof T. and Kessel, Pan and Gastegger, Michael and Nicoli, Kim A. and Tkatchenko, Alexandre and Müller, Klaus-Robert},
    title = "{SchNetPack: A Deep Learning Toolbox For Atomistic Systems}",
    journal = {J. Chem. Theory Comput.},
    volume = {15},
    number = {1},
    pages = {448-455},
    year = {2019},
    doi = {10.1021/acs.jctc.8b00908},
}

@article{ambrosetti2016wavelike,
  title={Wavelike charge density fluctuations and van der Waals interactions at the nanoscale},
  author={Ambrosetti, Alberto and Ferri, Nicola and DiStasio Jr, Robert A and Tkatchenko, Alexandre},
  journal={Science},
  volume={351},
  number={6278},
  pages={1171--1176},
  year={2016},
  publisher={American Association for the Advancement of Science},
  doi={10.1126/science.aae0509}
}

@article{wang2023aimd,
  title={AIMD-Chig: Exploring the conformational space of a 166-atom protein Chignolin with ab initio molecular dynamics},
  author={Wang, Tong and He, Xinheng and Li, Mingyu and Shao, Bin and Liu, Tie-Yan},
  journal={Sci. Data},
  volume={10},
  number={1},
  pages={549},
  year={2023},
  publisher={Nature Publishing Group UK London},
  doi={10.1038/s41597-023-02465-9}
}

@article{chmiela2023accurate,
  title={Accurate global machine learning force fields for molecules with hundreds of atoms},
  author={Chmiela, Stefan and Vassilev-Galindo, Valentin and Unke, Oliver T and Kabylda, Adil and Sauceda, Huziel E and Tkatchenko, Alexandre and M{\"u}ller, Klaus-Robert},
  journal={Sci. Adv.},
  volume={9},
  number={2},
  pages={eadf0873},
  year={2023},
  publisher={American Association for the Advancement of Science},
  doi={10.1126/sciadv.adf0873}
}

@article{hauseux2020quantum,
  title={From quantum to continuum mechanics in the delamination of atomically-thin layers from substrates},
  author={Hauseux, Paul and Nguyen, Thanh-Tung and Ambrosetti, Alberto and Ruiz, Katerine Saleme and Bordas, St{\'e}phane PA and Tkatchenko, Alexandre},
  journal={Nat. Commun.},
  volume={11},
  number={1},
  pages={1651},
  year={2020},
  publisher={Nature Publishing Group UK London},
  doi={10.1038/s41467-020-15480-w}
}

@article{hauseux2022colossal,
  title={Colossal enhancement of atomic force response in van der Waals materials arising from many-body electronic correlations},
  author={Hauseux, Paul and Ambrosetti, Alberto and Bordas, St{\'e}phane PA and Tkatchenko, Alexandre},
  journal={Phys. Rev. Lett.},
  volume={128},
  number={10},
  pages={106101},
  year={2022},
  publisher={APS},
  doi={10.1103/PhysRevLett.128.106101}
}

@article{schutt2021equivariant,
  title={Equivariant message passing for the prediction of tensorial properties and molecular spectra},
  author={Sch{\"u}tt, Kristof and Unke, Oliver and Gastegger, Michael},
  journal={Int. Conf. on Mach. Learn.},
  pages={9377--9388},
  volume={139},
  year={2021},
  url={https://proceedings.mlr.press/v139/schutt21a.html}
}

@book{bader1990atoms,
  title={Atoms in molecules: a quantum theory},
  author={Bader, Richard F. W.},
  year={1990},
  publisher={Oxford University Press},
  doi = {10.1093/oso/9780198551683.001.0001}
}

@article{blanco2005interacting,
  title={Interacting quantum atoms: a correlated energy decomposition scheme based on the quantum theory of atoms in molecules},
  author={Blanco, MA and Mart{\'\i}n Pend{\'a}s, A and Francisco, E},
  journal={J. Chem. Theory Comput.},
  volume={1},
  number={6},
  pages={1096--1109},
  year={2005},
  publisher={ACS Publications},
  doi={10.1021/ct0501093}
}

@article{bistoni2024local,
  title={Local energy decomposition analysis of London dispersion effects: From simple model dimers to complex biomolecular assemblies},
  author={Bistoni, Giovanni and Altun, Ahmet and Wang, Zikuan and Neese, Frank},
  journal={Acc. Chem. Res.},
  volume={57},
  number={9},
  pages={1411--1420},
  year={2024},
  publisher={ACS Publications},
  doi={10.1021/acs.accounts.4c00085}
}

@article{schreiber1996rapid,
  title={Rapid, electrostatically assisted association of proteins},
  author={Schreiber, Gideon and Fersht, Alan R},
  journal={Nat. Struct. Mol. Biol.},
  volume={3},
  number={5},
  pages={427--431},
  year={1996},
  publisher={Nature Publishing Group US New York},
  doi={10.1038/nsb0596-427}
}

@article{honda200410,
  title={10 residue folded peptide designed by segment statistics},
  author={Honda, Shinya and Yamasaki, Kazuhiko and Sawada, Yoshito and Morii, Hisayuki},
  journal={Structure},
  volume={12},
  number={8},
  pages={1507--1518},
  year={2004},
  publisher={Elsevier},
  doi={10.1016/j.str.2004.05.022}
}

@article{lindorff2011fast,
  title={How fast-folding proteins fold},
  author={Lindorff-Larsen, Kresten and Piana, Stefano and Dror, Ron O and Shaw, David E},
  journal={Science},
  volume={334},
  number={6055},
  pages={517--520},
  year={2011},
  publisher={American Association for the Advancement of Science},
  doi={10.1126/science.1208351}
}

@article{stohr2019quantum,
  title={Quantum mechanics of proteins in explicit water: The role of plasmon-like solute-solvent interactions},
  author={St{\"o}hr, Martin and Tkatchenko, Alexandre},
  journal={Sci. Adv.},
  volume={5},
  number={12},
  pages={eaax0024},
  year={2019},
  publisher={American Association for the Advancement of Science},
  doi={10.1126/sciadv.aax0024}
}

@article{jones2013quantum,
  title={Quantum Drude oscillator model of atoms and molecules: Many-body polarization and dispersion interactions for atomistic simulation},
  author={Jones, Andrew P and Crain, Jason and Sokhan, Vlad P and Whitfield, Troy W and Martyna, Glenn J},
  journal={Phys. Rev. B},
  volume={87},
  number={14},
  pages={144103},
  year={2013},
  publisher={APS},
  doi={10.1103/PhysRevB.87.144103}
}

@article{karimpour2026quantum,
  title={Quantum Field Approaches to Chemical Systems},
  author={Karimpour, Reza and Gori, Matteo and Tkatchenko, Alexandre},
  journal={arXiv preprint arXiv:2603.17582},
  year={2026},
  url={https://arxiv.org/abs/2603.17582}, 
}

@book{berazin2012method,
  title={The method of second quantization},
  author={Berezin, FA},
  year={1966},
  publisher={Academic Press}
}

@book{blaizot1986quantum,
  title={Quantum theory of finite systems},
  author={Blaizot, Jean-Paul and Ripka, Georges},
  publisher={MIT Press},
  year={1986}
}

@article{khabibrakhmanov2025accurate,
  title={Accurate noncovalent interactions in atomistic systems via quantum Drude oscillators},
  author={Khabibrakhmanov, Almaz and Fedorov, Dmitry V and Ambrosetti, Alberto and Crain, Jason and Hunt, Katharine LC and Johnson, Erin R and Jordan, Kenneth D and G{\'o}ger, Szabolcs and Gori, Matteo and Karimpour, Mohammad Reza and others},
  journal={J. Chem. Phys.},
  volume={163},
  number={15},
  pages={151001},
  year={2025},
  publisher={AIP Publishing},
  doi={10.1063/5.0281913}
}

@book{frenkel2023understanding,
  title={Understanding molecular simulation: from algorithms to applications},
  author={Frenkel, Daan and Smit, Berend},
  year={2023},
  publisher={Elsevier},
  url={https://doi.org/10.1016/C2009-0-63921-0}
}

@article{kabylda2025molecular,
  title={Molecular Simulations with a Pretrained Neural Network and Universal Pairwise Force Fields},
  author={Kabylda, Adil and Frank, J Thorben and Su{\'a}rez-Dou, Sergio and Khabibrakhmanov, Almaz and Medrano Sandonas, Leonardo and Unke, Oliver T and Chmiela, Stefan and M{\"u}ller, Klaus-Robert and Tkatchenko, Alexandre},
  journal={J. Am. Chem. Soc.},
  volume={147},
  number={37},
  pages={33723},
  year={2025},
  doi={10.1021/jacs.5c09558},
  url={https://doi.org/10.1021/jacs.5c09558}
}

@article{zhou2018electrostatic,
  title={Electrostatic interactions in protein structure, folding, binding, and condensation},
  author={Zhou, Huan-Xiang and Pang, Xiaodong},
  journal={Chem. Rev.},
  volume={118},
  number={4},
  pages={1691--1741},
  year={2018},
  publisher={ACS Publications},
  doi={10.1021/acs.chemrev.7b00305}
}

@article{cornell1995second,
  title={A second generation force field for the simulation of proteins, nucleic acids, and organic molecules},
  author={Cornell, Wendy D and Cieplak, Piotr and Bayly, Christopher I and Gould, Ian R and Merz, Kenneth M and Ferguson, David M and Spellmeyer, David C and Fox, Thomas and Caldwell, James W and Kollman, Peter A},
  journal={J. Am. Chem. Soc.},
  volume={117},
  number={19},
  pages={5179--5197},
  year={1995},
  publisher={ACS Publications},
  doi={10.1021/ja00124a002}
}

@article{mackerell1998all,
  title={All-atom empirical potential for molecular modeling and dynamics studies of proteins},
  author={MacKerell Jr, Alex D and Bashford, Donald and Bellott, MLDR and Dunbrack Jr, Roland Leslie and Evanseck, Jeffrey D and Field, Martin J and Fischer, Stefan and Gao, Jiali and Guo, Houyang and Ha, Sookhee and others},
  journal={J. Phys. Chem. B.},
  volume={102},
  number={18},
  pages={3586--3616},
  year={1998},
  publisher={ACS Publications},
  doi={10.1021/jp973084f}
}

@book{stone2013theory,
  title={The theory of intermolecular forces},
  author={Stone, Anthony},
  year={2013},
  publisher={Oxford University Press},
  url={https://doi.org/10.1093/acprof:oso/9780199672394.001.0001}
}

@article{hermann2017first,
  title={First-principles models for van der Waals interactions in molecules and materials: Concepts, theory, and applications},
  author={Hermann, Jan and DiStasio Jr, Robert A and Tkatchenko, Alexandre},
  journal={Chem. Rev.},
  volume={117},
  number={6},
  pages={4714--4758},
  year={2017},
  publisher={ACS Publications},
  doi={acs.chemrev.6b00446}
}

@article{stone2007atom,
  title={Atom--atom potentials from ab initio calculations},
  author={Stone, AJ and Misquitta, AJ},
  journal={Int. Rev. Phys. Chem.},
  volume={26},
  number={1},
  pages={193--222},
  year={2007},
  publisher={Taylor \& Francis},
  doi={10.1080/01442350601081931}
}

@article{london1937general,
  title={The general theory of molecular forces},
  author={London, Fritz},
  journal={Trans. Faraday Soc.},
  volume={33},
  pages={8b--26},
  year={1937},
  publisher={Royal Society of Chemistry},
  doi={10.1039/TF937330008B}
}

@article{jones1924determination,
  title={On the determination of molecular fields.—II. From the equation of state of a gas},
  author={Jones, John Edward},
  journal={Proc. R. Soc. A.},
  volume={106},
  number={738},
  pages={463--477},
  year={1924},
  publisher={The Royal Society London},
  doi={10.1098/rspa.1924.0082}
}

@article{grimme2010consistent,
  title={A consistent and accurate ab initio parametrization of density functional dispersion correction (DFT-D) for the 94 elements H-Pu},
  author={Grimme, Stefan and Antony, Jens and Ehrlich, Stephan and Krieg, Helge},
  journal={J. Chem. Phys.},
  volume={132},
  number={15},
  pages={154104},
  year={2010},
  publisher={AIP Publishing},
  doi={10.1063/1.3382344}
}

@article{tkatchenko2009accurate,
  title={Accurate molecular van der waals interactions from ground-state electron density and free-atom reference data},
  author={Tkatchenko, Alexandre and Scheffler, Matthias},
  journal={Phys. Rev. Lett.},
  volume={102},
  number={7},
  pages={073005},
  year={2009},
  publisher={APS},
  doi={10.1103/PhysRevLett.102.073005}
}

@article{blum2009ab,
  title={Ab initio molecular simulations with numeric atom-centered orbitals},
  author={Blum, Volker and Gehrke, Ralf and Hanke, Felix and Havu, Paula and Havu, Ville and Ren, Xinguo and Reuter, Karsten and Scheffler, Matthias},
  journal={Comput. Phys. Commun.},
  volume={180},
  number={11},
  pages={2175--2196},
  year={2009},
  publisher={Elsevier},
  doi={10.1016/j.cpc.2009.06.022}
}

@article{ren2012resolution,
  title={Resolution-of-identity approach to Hartree--Fock, hybrid density functionals, RPA, MP2 and GW with numeric atom-centered orbital basis functions},
  author={Ren, Xinguo and Rinke, Patrick and Blum, Volker and Wieferink, J{\"u}rgen and Tkatchenko, Alexandre and Sanfilippo, Andrea and Reuter, Karsten and Scheffler, Matthias},
  journal={New J. Phys.},
  volume={14},
  number={5},
  pages={053020},
  year={2012},
  publisher={IOP Publishing},
  doi={10.1088/1367-2630/14/5/053020}
}

@article{honig1995classical,
  title={Classical electrostatics in biology and chemistry},
  author={Honig, Barry and Nicholls, Anthony},
  journal={Science},
  volume={268},
  number={5214},
  pages={1144--1149},
  year={1995},
  publisher={American Association for the Advancement of Science},
  doi={10.1126/science.7761829}
}

@article{ren2012biomolecular,
  title={Biomolecular electrostatics and solvation: a computational perspective},
  author={Ren, Pengyu and Chun, Jaehun and Thomas, Dennis G and Schnieders, Michael J and Marucho, Marcelo and Zhang, Jiajing and Baker, Nathan A},
  journal={Q. Rev. Biophys.},
  volume={45},
  number={4},
  pages={427--491},
  year={2012},
  publisher={Cambridge University Press},
  doi={10.1017/S003358351200011X}
}

@article{motlagh2014ensemble,
  title={The ensemble nature of allostery},
  author={Motlagh, Hesam N and Wrabl, James O and Li, Jing and Hilser, Vincent J},
  journal={Nature},
  volume={508},
  number={7496},
  pages={331--339},
  year={2014},
  publisher={Nature Publishing Group UK London}
}

@article{piana2011computational,
  title={Computational design and experimental testing of the fastest-folding $\beta$-sheet protein},
  author={Piana, Stefano and Sarkar, Krishnarjun and Lindorff-Larsen, Kresten and Guo, Minghao and Gruebele, Martin and Shaw, David E},
  journal={J. Mol. Biol.},
  volume={405},
  number={1},
  pages={43--48},
  year={2011},
  publisher={Elsevier},
  doi={10.1016/j.jmb.2010.10.023}
}

@article{szalewicz2012symmetry,
  title={Symmetry-adapted perturbation theory of intermolecular forces},
  author={Szalewicz, Krzysztof},
  journal={WIREs Comput. Mol. Sci.},
  volume={2},
  number={2},
  pages={254--272},
  year={2012},
  publisher={Wiley Online Library},
  doi={10.1002/wcms.86}
}

@article{sali1994how,
  author  = {{\v{S}}ali, Andrej and Shakhnovich, Eugene and Karplus, Martin},
  title   = {How does a protein fold?},
  journal = {Nature},
  year    = {1994},
  volume  = {369},
  number  = {6477},
  pages   = {248--251},
  doi     = {10.1038/369248a0}
}

\clearpage
\renewcommand{\thesection}{S\arabic{section}}  
\renewcommand{\thetable}{S\arabic{table}}  
\renewcommand{\thefigure}{S\arabic{figure}}
\setcounter{figure}{0}
\setcounter{table}{0}
\onecolumngrid

\section{Supporting Information}


\begin{table*}[h]
  \caption{Dataset statistics and model errors for MBD reference data. Energy and force ranges refer to the MBD dispersion correction. Force range is the min-to-max range of Cartesian force components across all structures.}
  \centering
  \setlength{\tabcolsep}{6pt}
  \begin{tabular}{l c c c c c}
    \toprule
    \textbf{System}
      & \textbf{Model}
      & \makecell{\textbf{Energy MAE}\\(kcal/mol)}
      & \makecell{\textbf{Force MAE}\\(kcal/mol/\AA)}
      & \makecell{\textbf{Energy range}\\(kcal/mol)}
      & \makecell{\textbf{Force range}\\(kcal/mol/\AA)} \\
    \midrule
    \multirow{2}{*}{\ce{AcAla3NMe}}
      & MLFF & 0.040 & 0.008 & \multirow{2}{*}{6.5}  & \multirow{2}{*}{2.2} \\
      & MEFF & 0.770 & 0.510 &                       &                      \\
    \midrule
    \multirow{2}{*}{DHA}
      & MLFF & 0.040 & 0.009 & \multirow{2}{*}{12.1} & \multirow{2}{*}{3.8} \\
      & MEFF & 1.620 & 0.700 &                       &                      \\
    \midrule
    \multirow{2}{*}{\makecell[l]{Buckyball\\Catcher}}
      & MLFF & 0.070 & 0.010 & \multirow{2}{*}{18.4} & \multirow{2}{*}{3.4} \\
      & MEFF & 3.300 & 0.840 &                       &                      \\
    \midrule
    \multirow{2}{*}{Chignolin}
      & MLFF & 0.130 & 0.020 & \multirow{2}{*}{43.5} & \multirow{2}{*}{3.3} \\
      & MEFF & 3.910 & 0.530 &                       &                      \\
    \bottomrule
  \end{tabular}
  \label{tab:model_accuracy_mbd_extended}
\end{table*}

\begin{table*}[h]
  \caption{Dataset statistics and model errors for PBE0+MBD total energies and forces. Energy and force ranges refer to total PBE0+MBD values. Force range is the min-to-max range of Cartesian force components across all structures.}
  \centering
  \setlength{\tabcolsep}{6pt}
  \begin{tabular}{l l c c c c}
    \toprule
    \textbf{System}
      & \textbf{Model}
      & \makecell{\textbf{Energy MAE}\\(kcal/mol)}
      & \makecell{\textbf{Force MAE}\\(kcal/mol/\AA)}
      & \makecell{\textbf{E range}\\(kcal/mol)}
      & \makecell{\textbf{F range}\\(kcal/mol/\AA)} \\
    \midrule
    \multirow{2}{*}{\ce{AcAla3NMe}}
      & MLFF & 0.29 & 0.62 & \multirow{2}{*}{102}  & \multirow{2}{*}{438} \\
      & MEFF & 6.35 & 17.8 &                       &                      \\
    \midrule
    \multirow{2}{*}{DHA}
      & MLFF & 0.40 & 0.50 & \multirow{2}{*}{75.7} & \multirow{2}{*}{420} \\
      & MEFF & 8.15 & 18.4 &                       &                      \\
    \midrule
    \multirow{2}{*}{\makecell[l]{Buckyball\\Catcher}}
      & MLFF & 0.28 & 0.24 & \multirow{2}{*}{237}  & \multirow{2}{*}{341} \\
      & MEFF & 11.5 & 18.4 &                       &                      \\
    \midrule
    \multirow{2}{*}{Chignolin}
      & MLFF & 0.35 & 0.26 & \multirow{2}{*}{341}  & \multirow{2}{*}{480} \\
      & MEFF & 54.4 & 14.2 &                       &                      \\
    \bottomrule
  \end{tabular}
  \label{tab:painn_full_table}
\end{table*}

\begin{figure*}[h]
    \centering
    \includegraphics[width=.5\linewidth]{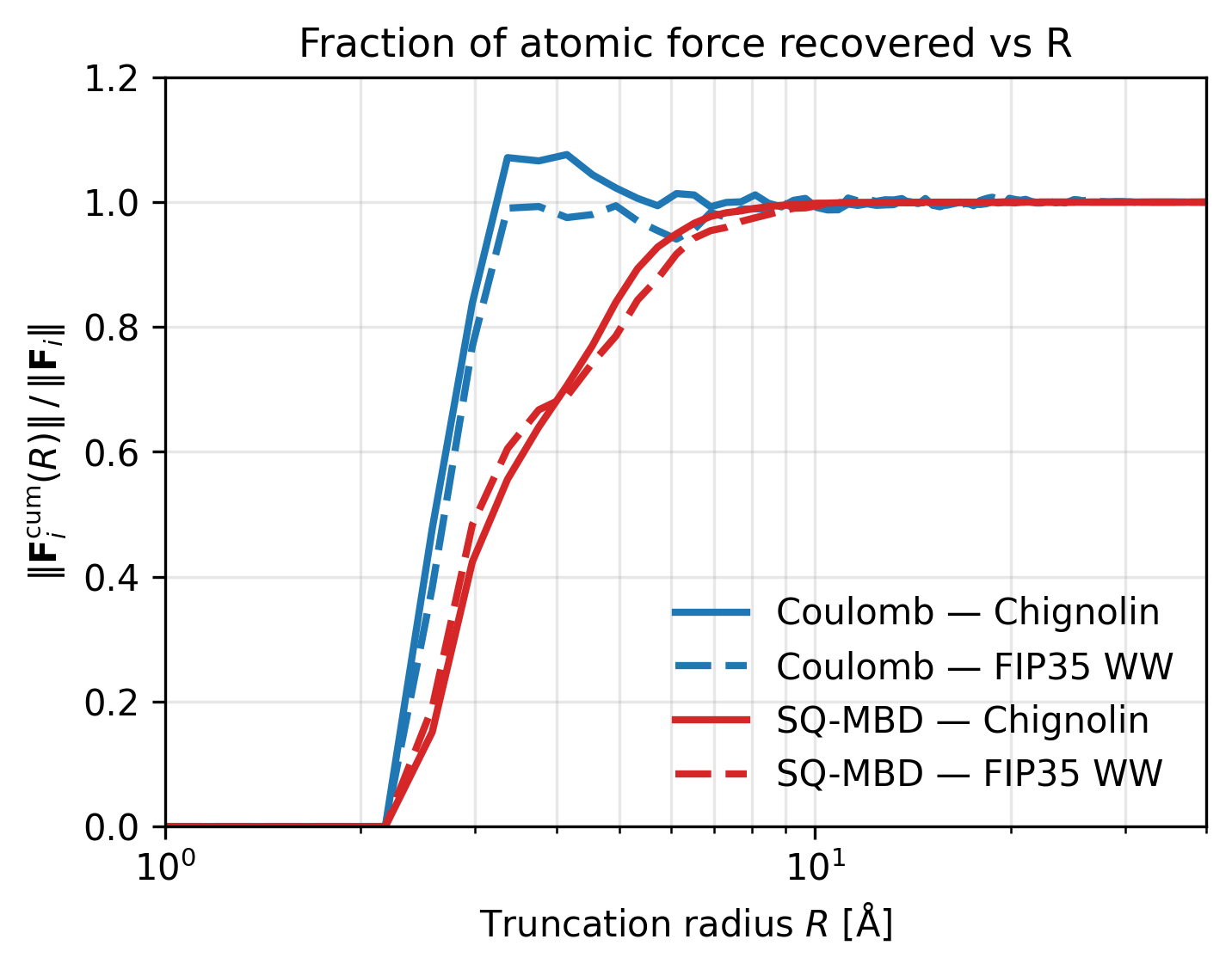}
    \caption{\textbf{Convergence of non-bonded atomic forces with truncation radius.} Fraction of the full non-bonded atomic-force magnitude recovered by truncating the pair sum at radius $R$, $\lVert\mathbf{F}_i^{\mathrm{cum}}(R)\rVert / \lVert\mathbf{F}_i\rVert$, for Chignolin (solid lines) and the FIP35 WW domain (dashed lines). Per-atom cumulative forces are computed for two (effective) pairwise components: point-charge Coulomb forces using SO3LR atomic partial charges (blue) and SQ-MBD pair forces obtained from the rsSCS pair-force matrix $F^{\mathrm{MBD}}_{ij}$ (red). Bonded 1–2 and 1–3 pairs are excluded from both sums (i.e. atoms separated by one or two covalent bonds, with bonds assigned by a 1.8~\AA{} distance cutoff). Curves show the median across atoms and trajectory frames. SQ-MBD forces converge more slowly than electrostatic forces with increasing truncation radius, reflecting the longer correlation range of the many-body dispersion response.}
    \label{fig:elec_mbd_conv}
\end{figure*}

\begin{figure*}
    \centering
    \includegraphics[width=.99\linewidth]{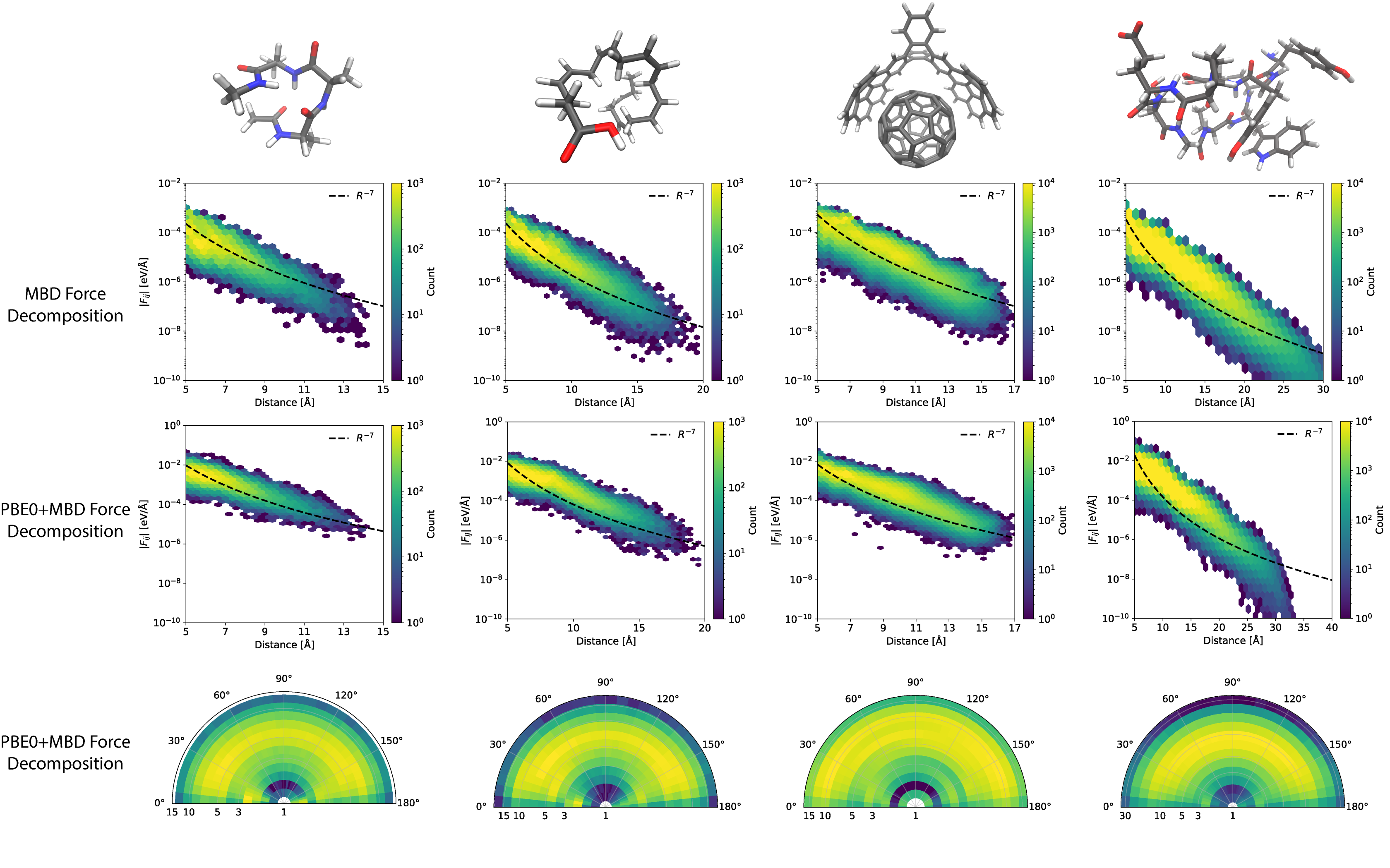}
    \caption{\textbf{PaiNN force decomposition: MBD versus PBE0+MBD.} Comparison of pairwise force decompositions from PaiNN MLFFs trained on two different reference targets, demonstrating that the interaction patterns reported in the main text are not specific to the long-range dispersion component. Columns correspond to \ce{AcAla3NMe} (42 atoms), DHA (56 atoms), the Buckyball Catcher (148 atoms), and Chignolin (166 atoms). First row: representative molecular geometries. Second row: interaction strength $|F_{ij}|$ versus interatomic distance for PaiNN trained on MBD-only forces, reproduced from Fig.~\ref{fig:magnitude} for direct comparison; dashed lines indicate $R^{-7}$ scaling. Third row: same decomposition for PaiNN trained on full PBE0+MBD forces, recovering the broad scatter observed in the MBD-only case. In the second and third rows, the dashed $R^{-7}$ reference is a least-squares fit of $|F_{ij}| = c\,R^{-7}$ to all samples in the 5--10~\AA{} range, pooled across element pairs and frames. Fourth row: radial heatmaps of the angle between $\mathbf{F}_{ij}$ and the interatomic displacement vector for the PBE0+MBD model ($180^\circ$ denotes attractive pairwise alignment). Heatmap colors encode bin counts on a logarithmic scale. Training statistics are reported in Table~\ref{tab:painn_full_table}.}
    \label{fig:painn_full}
\end{figure*}

\end{document}